\documentstyle[12pt]{article}

\def\abstract#1{\vskip 7mm 
	\begin{center}{\large Abstract}\par \bigskip
		\begin{minipage}[c]{12cm}
			\small #1
		\end{minipage}
	\end{center}
}
\def\title#1{\begin{center}{\Large\bf #1}\end{center}}
\def\author#1{\vskip 5mm \begin{center}{#1}\end{center}}
\def\address#1{\begin{center}{\it #1}\end{center}}

\newcommand{\bfr}{\begin{flushright}}
\newcommand{\efr}{\end{flushright}}

\begin{document}

\vspace*{-2cm}
\bfr{}\efr\vspace{-9mm}
\bfr{}\efr
\vspace{1cm}

\title{Discrete and Continuous Bogomolny Equations through the Deformed Algebra}\author{Takao Koikawa\footnote{e-mail address: koikawa@otsuma.ac.jp}}
\vspace{1cm}
\address{
  School of Social Information Studies\\
         Otsuma Women's University\\
         Tama, Tokyo 206, Japan\\
}
\vspace{2.5cm}
\abstract{ }
We connect the discrete and continuous Bogomolny equations. There exists one-parameter algebra relating two equations which is the deformation of the extended conformal algebra. This shows that the deformed algebra plays the role of the link between the matrix valued model and the model with one more space dimension higher.

\newpage
Lately the formulation of the 4-dimensional self-dual gravity called the heavenly equation\cite{Hus}-\cite{Cas} reveals an interesting connection with the 2-dimensional self dual Yang-Mills theory which may be formulated by the Matrix valued fields satisfying the zero curvature equation and the continuity equation. The heavenly equations are obtained from the self dual Yang-Mills theory by replacing the commutation relation by the Poisson bracket. In the connection the Moyal bracket plays an important role. It has a parameter and its adjustment connects the 4-dimensional self-dual gravity and the 2-dimensional self dual Yang-Mills theory. We may carry over 2-dimensional solution to the 4-dimensional solution by making use of the connection. Here we should note that the Poisson bracket emerges as the consequence of the parameter $\kappa \to 0$ limit in the Moyal algebra\cite{Moy}-\cite{Str}.

Also many people are now  studying the Matrix model of which the large N limit is the string model\cite{IKKT,BFSS}. By taking the $N \to \infty$ limit in the Matrix valued model we get the continuous action with even number of dimensions. Also in this field the Poisson bracket emerges and it is essential in formulating the geometrical model.

In these recent work of creating continuous space-time from discrete space, we always encounter the creation of even number of space-time dimensions. For example, we can add two dimensional space to the original 2-dimensional Matrix valued field equations by replacing the $su(2)$ Lie algebra commutation relations with the Poisson bracket. And successive application of the method might create the even number of new continuous dimensions by adopting the direct product of the algebra. But how can we obtain the odd number of new continuous dimension? This motivates the present work. We study how we get the equation with one dimension higher than the original Matrix valued equation. We also discuss the algebra behind its correspondence. Starting with the Bogomolny equation\cite{Bog} which is known to be identical to other nonlinear equations such as the nonlinear $\sigma$ model on the symmetric space\cite{CK}, we connect it with the continuos Bogomoly equation. Here one space is added to the original Matrix equation. In the connection the brackets isomorphic to the extended conformal algebra plays an important role of intervening the discrete and continuous dimension.

Construction of the present paper is as follows. We first introduce the extended conformal algebra and its isomorphic brackets. We then deform them. Defining the notations of the Chevalley basis, we show the Bogomolny equations expanded by them. We express the equations in terms of the deformed algebra with a fixed parameter. And then we take the limit of the parameter to get the continuous Bogomolny equations.

We define the operator $L_f$ by
\begin{equation}
L_f=-\frac{df(x)}{dx}+f(x) \frac{d}{dx},
\end{equation}
then when it is operated on a function $g(x)$ we get
\begin{equation}
	L_fg(x)=f(x)g'(x)-f'(x)g(x).
\end{equation}
Here the $f'(x)$ is $df(x)/dx$. We denote the rhs of this equation by the bracket $\{f(x),g(x)\}$. We can then show that the commutation relation between two operators satisfy
\begin{equation}
	[L_f,L_g]=L_{\{f,g\}}.
\end{equation}
Note that this is decomposed as
\begin{eqnarray}
	[f(x)\frac{d}{dx},g(x)\frac{d}{dx}] &=& \{f(x)g'(x)-f'(x)g(x)\}\frac{d}{dx},\\
	{[\frac{df(x)}{dx},g(x)\frac{d}{dx}]} &=& -\frac{d^2f(x)}{dx^2}g(x),\\
	{[\frac{df(x)}{dx},\frac{dg(x)}{dx}]} &=& 0.
\end{eqnarray}
The first commutation relation represents the conformal algebra. Usually we neglect the second and third commutation relations regarding them irrelevant to the algebra. But here we take them into account and consider their role. In the following we start with three sets of brackets isomorphic to these commutation relations:
\begin{eqnarray}
	\{f(x),g(x)\}_{11} &=& f(x)g'(x)-f'(x)g(x),\\
	\{\frac{d\psi(x)}{dx},g(x)\}_{01} &=& -\frac{d^2\psi(x)}{dx^2}g(x),\\
	\{\frac{d\psi(x)}{dx},\frac{d\phi(x)}{dx}\}_{00}&=&0,
\end{eqnarray}
where we assume that the range of $x$ is positive. We extend the brackets so that it may apply not only to $g(x)$ but also to $g(-x)$. In order to incorporate it we define the first and second bracket as
\begin{eqnarray}
	\{f(x),g(\pm x)\}_{11} &=&[f(x)\frac{dg(y)}{dx}-g(y)\frac{f(x)}{dy}]|_{y =\pm x},\\
	\{\frac{d\psi(x)}{dx},g(\pm x)\}_{01} &=& -[g(y) \frac{d}{dy}\left(\frac{d\psi(x)}{dx}\right)]|_{y =\pm x},
\end{eqnarray}
where the suffix $y = \pm x$ in the above equations is meant by the substitution before the derivative is taken . We thus obtain the following bracket relations;
\begin{eqnarray}
	\{f(x),g(-x)\}_{11} &=& \frac{d}{dx}(f(x)g(-x)),\\
	\{\frac{d\psi(x)}{dx},g(-x)\}_{01} &=&  \frac{d^2 \psi(x)}{dx^2}g(-x).
\end{eqnarray}
We now deform the brackets. In obtaining the Moyal bracket by deforming the Poisson bracket, many people have invented their own way to implement it. Strachan introduced the star product\cite{Str} which is the exponentiation of the Poisson bracket generators and obtained the Moyal algebra. We follow the method. We introduce the deformation in order to consider the finite displacement while undeformed brackets represent the algebra among the infinitesimal displacement generators. 
\begin{eqnarray}
	 f(x)\frac{dg(x)}{dx}-\frac{f(x)}{dx}g(x)&=& -Df(x) \circ g(x)\nonumber \\ 
	 &\to& -\frac{1}{\kappa}\rm{sh} \kappa D f(x) \circ g(x),
\end{eqnarray}
where the $D$ of $Df(x) \circ g(x)$ is the Hirota's bilinear operator on two functions $f(x)$ and $g(x)$, and $\kappa$ is a constant parameter.

Besides the deformations of the brackets, we should also deform the derivative of the first and second order. As for the exponentiations of the derivative operator to get a difference operator, there are two ways:the forward and backward differences given by
\begin{eqnarray}
	\frac{d}{dx}f(x) &\to& \frac{2}{\kappa}\rm{exp}(\frac{2}{\kappa}\frac{d}{dx})
	\rm{sh} \frac{2}{\kappa} \frac{d}{dx}f(x)=\triangle^* f(x),\\
	\frac{d}{dx}f(x) &\to& \frac{2}{\kappa}\rm{exp}(-\frac{2}{\kappa}\frac{d}{dx})
	\rm{sh} \frac{2}{\kappa} \frac{d}{dx}f(x)=\triangle f(x).
\end{eqnarray}
We shall choose the latter here for the first order derivative. We define the second order derivative by multiplying them:$\triangle^* \triangle$. To sum up, we get the following rules:
\begin{eqnarray}
	\frac{d}{dx}f(x) &\to& \triangle f(x),\\
	\frac{d^2}{dx^2}f(x) &\to& \triangle ^* \triangle f(x).
\end{eqnarray}
Applying the rules to the brackets, we obtain the deformed brackets which we shall denote with a prime:
\begin{eqnarray}
	\{f(x),g(x)\}'_{11} &=& -\frac{1}{\kappa}\rm{sh} \kappa D f(x) \circ g(x),\\
	\{f(x),g(-x)\}'_{11} &=& \triangle (f(x)g(-x)),\\
	\{\triangle \psi(x),g(x)\}'_{01} &=& -(\triangle^* \triangle \psi(x))g(x),\\
	\{\triangle \psi(x),g(-x)\}'_{01} &=& (\triangle^* \triangle \psi(x))g(-x),\\
	\{\triangle \psi(x),\triangle \phi(x)\}'_{00} &=&  0.	
\end{eqnarray}
To be explicit, these deformed brackets are rewritten as 
\begin{eqnarray}
	& &{}\{f(x),g(x)\}'_{11}  \nonumber \\
	&=& -\frac{1}{\kappa}[f(x+\kappa)g(x-\kappa)-f(x-\kappa)g(x+\kappa)],\\
	\label{eq:DBini}
	& &{}\{f(x),g(-x)\}'_{11} \nonumber \\ 
	&=& -\frac{1}{\kappa}[f(x)g(-x)-f(x-\kappa)g(-x+\kappa)],\\
	& &{}\{\frac{1}{\kappa}[\psi(x)-\psi(x-\kappa)],g(x)\}'_{01}  \nonumber \\
	&=& -\frac{1}{{\kappa}^2}[\psi(x+\kappa)-2\psi(x)+\psi(x-\kappa)]g(x),\\
	& &{}\{\frac{1}{\kappa}[\psi(x)-\psi(x-\kappa)],g(-x)\}'_{01}  \nonumber \\
	&=& \frac{1}{{\kappa}^2}[\psi(x+\kappa)-2\psi(x)+\psi(x-\kappa)]g(-x),\\
	& &{}\{\frac{1}{\kappa}[\psi(x)-\psi(x-\kappa)],\frac{1}{\kappa}[\phi(x)-\phi(x-\kappa)]\}'_{00} \nonumber \\ 
	&=&  0.
	\label{eq:DBfnl}
\end{eqnarray}
Note that the above deformed equations reduce to the undeformed brackets in the $\kappa \to 0$ limit.

Before we show the Bogomolny equation, we recapitulate our notations of the Lie algebra basis which would be used to expand our equations. The Chevalley basis is defined by
\begin{eqnarray}
	 {[H_{\bf \alpha},H_{\bf \beta}]} &=& 0,\\
	 {[H_{\bf \alpha},E_{\pm \bf \beta}]} &=& \pm E_{\pm \bf \beta}K_{\bf \alpha \beta},\\ 
	 {[E_{\bf \alpha},E_{- \bf \beta}]} &=& \pm H_{\bf \alpha} \delta_{\bf \alpha \beta},
\end{eqnarray}
where the vector suffices $\alpha$ and $\beta$ are simple roots and $K_{\bf \alpha \beta}$ is the Cartan matrix characterizing the algebra and it is defined by
\begin{equation}
K_{\bf \alpha \beta}=\frac{2\bf \alpha \cdot \beta}{|\bf \alpha||\bf \beta|}.
\end{equation}
The $SU(2)$ monopole can be generalized to larger group monopole by embedding $SU(2)$ into the group $G$. We shall discuss the Bogomolny equation with spherical symmetry, of which the rotational generators are the sum of ordinary angular momentum operators and the three generators of $G$ satisfying angular momentum algebra. By imposing the symmetry we can reduce the full algebra expanding the fields to the algebra constituting the Chevalley basis defined above.  We shall expand the Bogomolny equation by these basis. For the purpose we shall express the algebra by the matrices representing $SU(N+1)$. We shall adopt the integer numbers as a suffix to differentiate the matrices instead of using the root vectors. We represent the $SU(N+1)$ matrix basis by their non-zero components:
\begin{eqnarray}
	(H_i)_{n n} &=& \delta_{in}-\delta_{i n-1},\\
	(E_i)_{n n+1}&=& \delta_{in},\\
	(E_{-i})_{n+1 n}&=& \delta_{in},
\end{eqnarray}
where the suffices run over from 1 to N. These matrices satisfy the above definitions of the Chevalley basis with their root vectors replaced by the integer numbers and the Cartan matrix for $SU(N+1)$. We pay little attention to the end points of the suffix range because later on we shall take $N \to \infty$ limit and they get irrelevant in the discussion. 

	The Bogomolny equations\cite{Bog} with the spherical symmetry\cite{WG,LS} read
\begin{eqnarray}
	\frac{d\psi(r)}{dr} &=& \frac{1}{2}[N_{+}(r),N_{-}(r)],\\
	\frac{dN_{+}(r)}{dr} &=& [\psi(r),N_{+}(r)],\\
	\frac{dN_{-}(r)}{dr} &=& -[\psi(r),N_{-}(r)],
\end{eqnarray}
where $\psi(r)$ and $N_{\pm}(r)$ are expressed as
\begin{eqnarray}
	\psi(r) &=& \sum \psi_i (r)H_i,\\
	N_{+}(r) &=& \sum f_i(r)E_i,\\
	N_{-}(r) &=& \sum f^*_i(r)E_{-i}.
\end{eqnarray}
By use of the representations, we can obtain nonzero components of the above matrices: 
\begin{eqnarray}
	 (\sum \psi_i H_i)_{n n}=\psi_n(r)-\psi_{n-1}(r),\\
	 (\sum f_iE_i)_{n n+1}=f_n(r),\\
	 (\sum f^*_iE_{-i})_{n+1 n}=f^*_n(r).
\end{eqnarray}
Then nonzero components of the commutation relations on the Bogomolny equations are computed by using the definitions of the basis as
\begin{eqnarray}
	{\left([N_{+}(r),N_{-}(r)]\right)_{n n}}&=&\left([\sum f_iE_i,\sum f^*_iE_{-i}]\right)_{n n} \nonumber \\
	&=&|f_n(r)|^2-|f_{n-1}(r)|^2, \\
	{\left([\psi(r),N_{+}(r)]\right)}_{n n+1}&=&\left([\sum \psi_i H_i,\sum f_iE_{i}]\right)_{n n+1} \nonumber\\ 
	&=& f_n(r)(-\psi_{n-1}(r)+2\psi_{n}(r)-\psi_{n+1}(r)),\\ 
	{\left([\psi(r),N_{-}(r)]\right)}_{n+1 n}&=&\left([\sum \psi_i H_i,\sum f^*_iE_{-i}]\right)_{n+1 n} \nonumber \\ 
	&=& -f^*_n(r)(-\psi_{n-1}(r)+2\psi_{n}(r)-\psi_{n+1}(r)).
\end{eqnarray}
We can now write down the nonzero components of the Bogomolny equations as
\begin{eqnarray}
	\frac{d}{dr}(\psi_n(r)-\psi_{n-1}(r)) &=& \frac{1}{2}(|f_n(r)|^2-|f_{n-1}(r)|^2),\\
	\label{eq:DBOGOini}
	{\frac{df_n(r)}{dr}} &=& f_n(r)(-\psi_{n-1}(r)+2\psi_{n}(r)-\psi_{n+1}(r)),\\
	{\frac{df^*_n(r)}{dr}} &=& f^*_n(r)(-\psi_{n-1}(r)+2\psi_{n}(r)-\psi_{n+1}(r)).
	\label{eq:DBOGOfnl}
\end{eqnarray}
Introducing $\rho_n(r)$ by
\begin{equation}
	\rho_n(r)=-\rm{ln}|f_n|^2,
\end{equation}
we obtain from the above equations
\begin{equation}
	\frac{d^2 \rho_n(r)}{dr^2}=-\rm{e}^{-\rho_{n-1}}+2\rm{e}^{-\rho_{n}}-\rm{e}^{-\rho_{n+1}},
\end{equation}
which is identical to the Toda lattice equation with imaginary time $r$. 

We shall relate the Bogomolny equations defined on the one dimensional lattice space originated from the root space and that on the continuous space through the deformed algebra defined in Eqs.(\ref{eq:DBini})-(\ref{eq:DBfnl}). 
We shall call the equations the discrete Bogomolny equations and the continuous Bogomoly equations, respectively. The Toda lattice is the completely integrable system and so is the discrete Bogomolny equation, which means that we can construct the exact solutions. 
Once we establish the appropriate relationship between the discrete and continuous Bogomolny equations, the solutions to the discrete Bogomolny equations are carried over to those to the continuous Bogomolny equations despite that one dimension is added there. 
The bridge relating the discrete and continuous space is the deformed algebra. 
Each field has an index which takes the integer numbers and it would be modified to the argument of field in the continuous model. 
Note that, in the deformed algebra, the argument has the discrete structure with a space $\kappa$ as seen in Eqs.(\ref{eq:DBini})-(\ref{eq:DBfnl}). 
By requiring that the continuous variable $x$ be integers and $\kappa$ be one, we can regard them as the index in the discrete model. 
We shall rewrite the above Eqs.(\ref{eq:DBOGOini})-(\ref{eq:DBOGOfnl}) by using the deformed algebra. Note that the lhs of the Bogomolny equations might be written as
\begin{eqnarray}
	\psi_n(r)-\psi_{n-1}(r)&\sim&[\psi(r;n\kappa)-\psi(r;(n-1)\kappa)]/\kappa|_{\kappa=1}, \nonumber \\
	&=&[\psi(r;x)-\psi(r;x-\kappa)]/\kappa,\\
	f_n(r)&\sim& f(r;n\kappa)|_{\kappa=1}=f(r;x),\\
	f^*_n(r)&\sim& f^*(r;n\kappa)|_{\kappa=1}=f^*(r;x), 
\end{eqnarray}
where $\kappa=1$ is assumed. So far the suffix $n$ of $f_n(r)$ is assumed to take positive integer values, but now we define $f_{-n}(r)$by $f_{-n}(r)=f_n(r)$ for positive integer n , which leads to $f(r;-x)=f^*(r;x)$.
Comparing the rhs of the Bogomolny equations and the deformed brackets in Eqs.(\ref{eq:DBini})-(\ref{eq:DBfnl}), we might write the Bogomolny equations in terms of the deformed brackets with the fields with two variables $x$ and $r$. Finally we reduce the discrete Bogomolny equations to
\begin{eqnarray}
	\frac{\partial }{\partial r}\left(\frac{\psi(r;x)-\psi(r;x-\kappa)}{\kappa}\right) 
	&=& \frac{1}{2}\{f(r;x),f^*(r;x)\}'_{11}\\
	{\frac{\partial f(r;x)}{\partial r}} &=& \{\frac{\psi(r;x)-\psi(r;x-\kappa)}{\kappa},f(r;x)\}'_{01},\\
	{\frac{\partial f^*(r;x)}{\partial r}} &=& \{\frac{\psi(r;x)-\psi(r;x-\kappa)}{\kappa},f^*(r;x)\}'_{01},
\end{eqnarray}
where $x \in Z$  and $\kappa=1$. In the above equations $\kappa$ is fixed to be one in order to be identical to the discrete Bogomolny equation, but from now on we regard it arbitrary and let it approach to zero so that we get the continuous limit of these equations. In order to implement this, we need to replace the fields on the lattice by that on the continuous space. In the discrete Toda lattice equation, $\kappa$ is the lattice spacing. We assume that $l$ is the total length of the one dimensional lattice and $N$ particles are sitting with the equal spacing $\kappa$, which leads to $\kappa=l/(N-1)$. With this assignment, the position $x$ on the lattice is connected with the integer index $n$ as $x=n\kappa(n=0,1,\dots,N-1)$. We take the $N \to \infty$ limit first and then take the $l \to \infty$ limit. In the first limit, $\kappa \to 0$ and so the above equations become the continuous equations. The range of $x$ which at this stage is $(0,l)$ becomes $(0,\infty)$ by the succeeding limit. The range of $x$ can be tuned as $(-\infty, \infty)$ by the change of variable. We thus connect the discrete and continuous Bogomolny equations. 

We shall see the detail in order. Let us take the $\kappa \to 0$ limit of the nonzero components of the matrix components:
\begin{eqnarray}
	[\psi(r;n\kappa)-\psi(r;(n-1)\kappa)]/\kappa 
	&\to& \frac{\partial \psi(r;x)}{\partial x}, \\
	f(r,n\kappa)  &\to& f(r;x),\\
	f^*(r,n\kappa) &\to& f^*(r;x),\\ 
	\quad \rm{as} \quad \kappa &\to& 0.\nonumber
\end{eqnarray}
At the same time, the deformed brackets are replaced by the non-deformed brackets:
\begin{eqnarray}
	\{f(r;x),f^*(r;x)\}' &\to& \{f(r;x),f^*(r;x)\},\\
	\{\frac{1}{\kappa}[\psi(x)-\psi(x-\kappa)],f(r;x)\}' &\to& \{\frac{\partial \psi(r;x)}	  {\partial x},f(r;x)\},\\ 
	\{\frac{1}{\kappa}[\psi(x)-\psi(x-\kappa)],f^*(r;x)\}'&\to& \{\frac
	{\partial \psi(r;x)}{\partial x},f^*(r;x)\},\\ \quad \rm{as} \quad \kappa &\to& 0.
	\nonumber
\end{eqnarray}
Using these limits, we obtain the continuous Bogomolny equations:
\begin{eqnarray}
	\frac{\partial^2 \psi(r;x)}{\partial r \partial x} &=& \frac{1}{2}\{f(r;x),f^*(r;x)\}_{11},\\
	{\frac{\partial f(r;x)}{\partial r}} &=& \{\frac{\partial \psi(r;x)}{\partial x},f(r;x)\}_{01},\\
	{\frac{\partial f^*(r;x)}{\partial r}} &=& -\{\frac{\partial \psi(r;x)}{\partial x},f^*(r;x)\}_{01},
\end{eqnarray}
or
\begin{eqnarray}
	\frac{\partial^2 \psi(r;x)}{\partial r \partial x} &=& \frac{1}{2}\frac{\partial}
	{\partial x}|f(r;x)|^2,\\
	{\frac{\partial f(r;x)}{\partial r}} &=& -\frac{\partial^2 \psi(r;x)}{\partial x^2}
	f(r;x),\\
	{\frac{\partial f^*(r;x)}{\partial r}} &=& -\frac{\partial^2 \psi(r;x)}{\partial x^2}
	f^*(r;x).
\end{eqnarray}
From these equations we can derive
\begin{eqnarray}
	\frac{\partial \psi(r;x)}{\partial r} &=& \frac{1}{2}|f(r;x)|^2,\\
	{\frac{\partial }{\partial r}\rm{ln}f(r;x)} &=& -\frac{\partial^2 \psi(r;x)}{\partial x^2},\\
	{\frac{\partial }{\partial r}\rm{ln}f^*(r;x)} &=& \frac{\partial^2 \psi(r;x)}{\partial x^2}.
\end{eqnarray}
Introducing $\rho (r;x)$ by
\begin{equation}
\frac{\partial}{\partial x}\rho (r;x)=- \rm{ln}|f(r;x)|^2,
\end{equation}
we can combine the set of Bogomolny equations as
\begin{equation}
	\frac{\partial ^2}{\partial r^2}\rho (r;x)=- \frac{\partial}{\partial x}
	\rm{e}^{-\frac{\partial}{\partial x}\rho(r;x)},
\end{equation}
which is the continuous Bogomolny equation or the continuous Toda lattice equation with an imaginary time $r$. In ref.\cite{Koi} we derived this equation and discussed its completely integrable character where we introduced $\rho(r;x)$ as a field implying the potential in the linear lattice. 

Alternatively we can derive other continuous Bogomolny equations.  Defining $\tilde \rho(r;x)$ by
\begin{equation}
\tilde\rho (r;x)=- \rm{ln}|f(r;x)|^2,
\end{equation}
we obtain
\begin{equation}
	\frac{\partial ^2}{\partial r^2}\tilde\rho (r;x)=- \frac{\partial^2}{\partial x^2}
	\rm{e}^{-\tilde \rho(r;x)}.
\end{equation}
This equation was discussed in ref.\cite{BF}. The relation between $\rho(r;x)$ and $\tilde \rho(r;x)$ is easy to find from the definitions:
\begin{equation}
\tilde \rho(r;x)= \frac{\partial}{\partial x} \rho(r;x).
\end{equation}

In summary, we start with the extension of the conformal algebra and obtained the deformed algebra which comprises the discrete Bogomolny equations when $\kappa=1$ and the continuous Bogomolny equation when $\kappa \to 0$. They are both completely integrable equations in one and two dimensions. This shows that the deformed algebra plays the role of the link between the matrix valued model and the model with one more space dimension higher.
\vspace{2cm}

\noindent
Acknowledgement

The author would like to acknowledge Prof. Saito for discussion.

\end{document}